\documentclass[aps,prl,nopacs,superscriptaddress]{revtex4-1}
\usepackage{amsmath}
\usepackage{epsfig}
\usepackage{amssymb}
\usepackage{dcolumn}
\usepackage{graphicx}
\usepackage{dcolumn}

\usepackage{color}

\newcommand{\beq}{\begin{equation}}
\newcommand{\eeq}{\end{equation}}

\newcommand{\pv}{\mathbf{p}}
\newcommand{\rv}{\mathbf{r}}
\newcommand{\vv}{\mathbf{v}}
\newcommand{\kv}{\mathbf{k}}
\newcommand{\Kv}{\mathbf{K}}

\begin{document}

\title{Multiple scattering induced negative refraction of matter waves}

\author{Florian Pinsker}
\email{florian.pinsker@gmail.com}

\affiliation{Department of Atomic and Laser Physics, University of Oxford, United Kingdom.}

\begin{abstract}

Starting from fundamental multiple scattering theory it is shown that negative refraction indices are feasible  for matter waves passing a well-defined ensemble of scatterers. A simple approach to this topic is presented and explicit examples for systems of scatterers in $1$D and $3$D are stated that imply negative refraction for a generic incoming quantum wave packet.  Essential features of the effective scattering field, densities and frequency spectrum of scatterers are considered. Additionally it is shown that negative refraction indices allow perfect transmission of the wave passing the ensemble of scatterers. Finally the concept of the superlens is discussed, since it is based on negative refraction and can be extended to matter waves utilizing the observations presented in this paper. This paves the way for `untouchable' quantum systems in analogy to cloaking devices for electromagnetic waves.

\end{abstract}

\pacs{03.65.Ca, 03.65.Nk, 03.65.Ta, 03.75.Be}

\maketitle

Negative refraction of electromagnetic (EM) waves was first discovered by V. G. Veselago in \cite{sovj} and in particular it has been noted that at the negative refractive material's surface waves entering the material are bent towards negative angles. Decades later meta-materials with this property were developed and they were found to give rise to the phenomenon of sub-wavelength localisation (superlensing). In addition to perfect localisation of the EM waves those materials imply complete transmission of the incoming wave passing through those so called superlenses \cite{P,P1,P2}. For a long time negative refraction was considered a mathematical curiosity rather than a feasible physical phenomenon shown in experiments, it is only due to the invention of artificial meta-materials that the necessary conditions could be provided experimentally \cite{P1,P2,nature} and by the means of those materials sub-wavelength localisation was shown \cite{Turk}. Due to the huge impact of sub-wavelength localisation of EM waves a natural question to ask is: Are negative refractive materials feasible for matter waves as well and what is necessary to refract matter waves in a similar way as EM waves? To give answers to those questions in this work I will utilize multiple scattering theory, which has been widely applied to derive refraction indices for matter waves \cite{R1,R2,R3,Lax}. It is worth noting that multiple quantum scattering theory is an application of the full quantum many-body problem for fixed particle numbers that allows a more practical viewpoint when single particle scattering on a cloud of potentially randomly positioned scatterers is of interest \cite{Lax}. The refraction indices $n$  derived so far by this approach are however approximations and their validity is particularly confined to the case  $n > 0$.  So one may ask: Are matter waves restricted to refract positively, perhaps due to a property of the governing many-body Schr\"odinger equation? In fact, this work discusses the feasibility of negative refraction in a way that is consistent with the quantum many-body problem and conditions for negative refraction are stated, i.e. scenarios of ensembles of scatterers implying negative refraction. Finally based on the concept of negative refraction the concept of superlenses \cite{P1,P2} for matter waves is discussed. Although the presented theory is new I point out that the topic of negative refraction for matter waves was considered before in \cite{Bou} and a specific proposal in terms of the single particle Schr\"odinger equation including external magnetic fields was regarded as a candidate for negative refraction, while the results presented in \cite{Bou} were found to be inaccurate \cite{Bouno}. In addition only recently Veselago-type lensing was successfully implemented utilizing ultra cold atoms confined to optical lattices by switching between the positive and negative energy branches of the many-body states \cite{lens}. In contrast to the proposal \cite{Bou} and the realisation \cite{lens} this work follows a different path showing that negative refraction is a property of certain closed quantum many-body systems (with fixed particle numbers) without the necessity to apply external fields, feasible only due to a proper arrangement and choice of scatterers.

\emph{Multiple scattering theory.--} The fundamental multiple scattering equations for quantum matter waves were first introduced by L. L. Foldy and extend the classical treatment based on the Boltzmann integro-differential equations when interference effects of the scattered particle become important \cite{Foldy}. The set of equations usually used to describe the scenario for point scatterers where the strength of the scattered wave from a scatterer is proportional to the external field acting on it  read as follows \cite{Foldy, Lax, Fundi,Forrey},
\beq\label{pa}
\Psi(\rv) = \Psi_0 \left( \rv \right) + \sum_i G(\rv - \rv_i) f_i \xi^i (\rv_i).
\eeq
Here the l.h.s. is the wave function generated by multiple scattering processes that are induced by a system of $N$ randomly arranged particles/obstacles and
\beq\label{xi}
\xi^i(\rv_i) = \Psi_0 \left( \rv_i \right) + \sum_{i\neq j} G(\rv_j - \rv_i) f_j \Psi(\rv_j)
\eeq
describes the local field which generates the scattered wave at positions $\rv_i$.  $\Psi_0 \left( \rv \right)$ is the incident  wave, i.e. the wave without the effects of scatterers, $\xi^i (\rv_i)$ is the local field at $\rv_i$ where the $i$th scatterer e.g. a noble gas atom is located at a fixed position -  importantly $\xi^i (\rv_i)$ can be approximated by $\xi_i = \exp(i \kv \cdot \rv_i)$ when all multiple scattering processes at scatterer $i$ are neglected \cite{Foldy}, i.e. only the first order scattering from each scatterer $i$ of the ensemble of multiple scatterers is considered. $f_i$ is the scattering amplitude/coefficient that relates the scattered wave to the scatterer and which in general is a function of the wave vector and position in space. In $3$D the behavior of the particle close to the $i$th scatterer is assumed to be \cite{Foldy,Lax,Fundi} $G(\rv ) = \exp\left(\pm i  |\kv| \cdot |\rv | \right)/(4 \pi |\rv |)$, which is the Green's function describing the propagation of the wave scattered from the $i$th particle between the scattering particles. This form assumes a spherically symmetric scattering process (s-wave scattering) \cite{Foldy}. The corresponding Green's function for a $2$D system is the Hankel function and for the $1$D system it is a plane wave, which will be introduced later. It should be noted that from \eqref{pa} and \eqref{xi} a plethora of derivations have been made particularly for refractive indices of matter waves and comparison with experiments show excellent agreement with data e.g. in experiments with monatomic sodium gases \cite{R1,R2,R3,Lax,Foldy,Fundi,Forrey}. 

\emph{Statement of the problem.--} Let us assume $\Psi_0$ and $\Psi$ to approximately resemble plane waves, which can be regarded a valid assumption for waves far away from the scattering processes and they form a basis for general localised wave packets. Given that and utilizing the mathematical notions from above we can state the scenario of interest as follows: For
\begin{equation}\label{form}
\sum_i G(\rv - \rv_i) f_i \xi^i (\rv_i) =  \exp\left(i \Kv \cdot \rv \right)- \exp\left( i \kv \cdot \rv \right)
\end{equation}
do amplitudes $f_i = f_i(\rv_i, \kv)$, local fields $\xi^i(\rv_i, \kv)$ and configurations of scatterers exist that extinguish the initial field and generate a new field which moves in opposite direction?  Thus formally we are interested in states where the incoming wave with wave vector $\kv$ implies an outgoing wave with wave vector $\Kv$ so that 
\beq\label{re}
\Kv = - \kv \equiv n \kv.
\eeq
This equation defines the refraction index for the matter wave $n$ \cite{Lax}, which by definition of the problem  is negative. Note that although at first sight negative refraction with $n=-1$ could be confused with the properties of a perfect mirror with reflection coefficient $R=1$, but in contrast waves refracted negatively occupy the space `behind' the scatterers as well. As shown later there even is no loss of amplitude of the scattered wave after passing the scatterers for certain stated parameters. The remainder of this work is dedicated to show which local fields and amplitudes are necessary to imply the refractive indices to be negative and of absolute value $1$ and further properties of those systems. The motivation to investigate the case of negative refraction stems from \cite{P}, where meta-materials allow sub-wavelength localisation for electromagnetic matter waves and the argumentation is based on the extinction theorem for matter waves \cite{Foldy}.  Mathematically the l.h.s. of \eqref{form} consists of a set of homogeneous linear algebraic equations and to make statements about $\xi^i, f_i$ etc.  approximations can be made. Furthermore as plane waves are simplifications we will also discuss wave packets, i.e. superpositions of plane waves, moving in opposite directions, such that for each component of the wave packet's spectrum \eqref{re} is satisfied.

\emph{Simplified state equation.--} Alternatively to solving eq. $\ref{pa}$ directly one can proceed as follows. As discussed in detail by \cite{Foldy} eq. $\ref{pa}$ can be simplified when considering the statistical average over all possible random positions of scatterers, which is our first assumption, while the case of fixed positions is trivially included. Then by further supposing the scatterers not to have internal degrees of freedom and to occupy a finite volume $V$ one effectively considers
\begin{equation}\label{ha}
\langle \Psi(\rv) \rangle = \exp\left(i \kv \cdot \rv \right) + \sum_i \int_V G(\rv - \rv_i) f_i \langle \xi^i (\rv_i) \rangle_i  d\rv_i.
\end{equation}
Here $\langle \xi^i (\rv_i) \rangle_i  $ is the scatterer induced field acting on the $i$th scatterer, which is averaged over all possible configurations of all the other scatterers \cite{Foldy}.  This field can be approximately replaced by an averaged field, which is a fair approximation when the number of scatterers $N$ is large \cite{Foldy} and supposing that we substitute $\langle \xi^i (\rv_i) \rangle_i  Ê\to \langle \xi (\rv_i) \rangle  $. So \eqref{ha} reduces to the simplified state equation
\begin{equation}\label{sol}
\langle \Psi(\rv) \rangle = \exp\left(i \kv \cdot \rv \right) + \int_V G(\rv - \rv') f (\rv')  \rho(\rv') \xi (\rv')  d\rv' \equiv \psi(\rv),
\end{equation}
where $\psi (\rv)$ is the coherent wave, i.e. the {\it average} matter wave function for all possible configurations of scatterers. $\xi(\rv)$ denotes the average local field induced by the scatterers and $f (\rv)$ the average amplitude - both functions  are given by \cite{Foldy, Fundi}
\beq
\xi(\rv) = \sum_i  \left \langle \xi_i \delta(\rv - \rv_i)  \right \rangle
\eeq
\beq
f(\rv) = \sum_i  \left \langle f_i \delta(\rv - \rv_i)  \right \rangle
\eeq
and $\rho(\rv)$ is the number density of scatterers. As $N \to \infty$ for scatterers close to each other compared to other length scales of the system, those sums can be approximated as well as integrals.

\emph{Positive refraction.--}
While this paper follows the aim to derive wave functions for negative refraction indices we give a short overview how positive refraction can be derived easily from \eqref{sol}. The macroscopic one-body Schr\"odinger equation equation for the average wave function is \cite{Fundi} 
\beq\label{masch}
- \frac{\hbar^2}{2m} \Delta \psi(\rv) + v(r) = E \psi(\rv),
\eeq
where $E$ is the incident particle energy and $m$ its mass. In terms of  eq. \ref{masch} \eqref{sol} is a solution for the optical potential
\beq\label{op}
v (\rv) = \begin{cases} - 2\pi \hbar^2 \rho  f \xi(\rv)/m  \hspace{14mm} \text{inside} \hspace{5mm}$V$\\ 0 \hspace{35mm} \text{outside} \hspace{5mm} $V$\end{cases},
\eeq
when assuming $\xi = c \psi$ i.e. to be proportional by a number $c$ and for the incident  wave energy given by $E = \hbar^2 k^2/(2m) $ \cite{Fundi}. From \eqref{op} one derives the simple relation $K'^2 = k^2 + 4 \pi \rho f c$  and thus a refraction index $n = K'/k$, with $K'$ denoting the wave vector of the generated wave. This derivation of the index of refraction, however  cannot address the case of negative refraction as only a formula for the squares of the wave vectors is given. In order to extend the discussion to negative refraction indices we may  proceed as follows.

\emph{Wave packets.--} To develop the formalism our starting point is a general scalar matter wave function describing a quantum particle in $D$ spatial dimensions
\beq\label{mw}
\psi(\rv,t) = \int_{\mathbb R^D} d \kv \cdot  g(\kv) e^{i (- \kv \cdot \rv - \omega t)},
\eeq
with $g(\kv)$ being the distribution function of different frequency components of the wave.
Correspondingly the de Broglie and Einstein relations which associate the wave properties with that of a massive `projectile'  are \cite{Brog} 
\beq\label{pv}
\pv = m \vv = \hbar \kv
\eeq 
\beq\label{E}
E = \hbar \omega = h \nu.
\eeq
Here $\pv$ denotes the momentum of the projectile, $\vv$ its velocity, $\nu$ the frequency and $E$ the energy of the particle to be scattered. We consider the pair of initial and outgoing wave formally  given by
\beq\label{mw2}
\psi(\rv)_{\rm in, \rm out} = \int_{\mathbb R^D} d \kv \cdot  g(\kv) e^{\pm i \kv \cdot \rv},
\eeq
when incoming $-$ and outgoing waves $+$ are stationary.
\[
\]

\section{Results}

\subsection{Negative refraction in $1$D}
We are interested in the scenario where the initial wave packet $\psi(\rv)_{\rm in}$ is moving in opposite direction as the outgoing wave packet $\psi(\rv)_{\rm out}$ and we first confine our considerations to $D=1$. To model two counter-propagating waves we consider by using eq. \ref{sol} the expression
\begin{equation}\label{12}
g(k) (e^{i k \cdot x} - e^{- i k \cdot x})   = \int_V G^{\rm 1D} (x- x') f (k,x')  \alpha^{\rm 1D} (x') d x'.
\end{equation}
The superscript of the Green's function indicates its spacial dimension. Here we have introduced the abbreviation for the unknown $ \alpha^{\rm 1D} (x')   =   \rho(x') \xi (x')$, which we determine as follows. Let us rewrite the r.h.s. of \eqref{12} as
\begin{equation}\label{ooo}
 \int_V G^{\rm 1D}(x - x') f (k,x')  \alpha^{\rm 1D} (x')  d x' =  \int^\infty_{-\infty} G^{\rm 1D}(x - x') f (k,x') \alpha^{\rm 1D} (x') d x', 
 \end{equation}
and assume $\alpha^{\rm 1D}(x')$ to be compactly supported on $V$, which is a fairly reasonable statement  in physical terms particularly as the density of scatterers is only supported in $V$. 
Now for a $1$D system the Green's function of the inhomogeneous Helmholtz equation
\beq
\partial^2_x G^{\rm 1D}(x) + k^2 G^{\rm 1D}(x) = -\delta(x) \text{ with } x \in \mathbb R \, 
\eeq
corresponding to isotropic point scatterers at $x=0$ is
\beq\label{green1d}
G^{\rm 1D}_{\pm}(x) = \frac{e^{\pm i |k| |x|}}{2 |k|},
\eeq
which provides the behaviour of the particle close to the scatterer with incident wave number $|k| = \sqrt{2m E}/\hbar$. Clearly linear combinations of Green's functions are Green's functions as well and so we obtain by the above solutions e.g.
\begin{equation}
G^{\rm 1D}_0(x) =  \left( \frac{e^{ i |k| |x|}}{2 |k|} +\frac{e^{- i |k| |x|}}{2 |k|} \right) =\frac{\cos (- |k| |x|)}{|k|},
\end{equation}
i.e. a stationary wave solution and similarly one gets $\frac{-i \sin (- i |k| |x|)}{|k|}$. 

By combining \eqref{12} and  \eqref{ooo} we obtain the condition for negative refraction in $1$D when writing the two complex exponentials as sine,
\begin{equation}\label{13}
\int^\infty_{-\infty}f (|k|) \cdot G^{\rm 1D} (x-x')   \cdot  \alpha^{\rm 1D} (x') d x'  \equiv 2 i g(k)\sin{(k \cdot x)}.
\end{equation}
We assume the amplitude $f$ to be independent of $x'$, i.e. the strength of the scattered wave is only proportional to the Green's function and the number density of scatterers. In principle scatterers that obey a position dependency of the scattered wave beyond that of the Green's function of each scatterer, i.e. in $f$ can be simplified by effectively choosing the number distribution of scatterers $\rho(x)$ accordingly. As $\alpha^{\rm 1D}(x)$ is a product of the density of scatterers $\rho (x)$ and the local effective field $\xi(x)$ there is in principle freedom to generate a variety of $\alpha^{\rm 1D}(x)$ experimentally by arranging scatterers appropriately, which fixes $\rho (x)$. It is thus the choice of the experimenter first to use scatterers that indeed obey a position independency in $f$ and secondly to effectively compensate this dependency by the arrangement of scatterers. Note that the nuclear scattering amplitude has an expansion of the form \cite{Fundi}
\beq
f(k) = - b_i + i k b^2_i + \mathcal O(k^2) 
\eeq
with the bound scattering length $b_i$.

\subsection{Examples for $1$D scattering}

\emph{ Example $1$.--}
In the following we will give explicit examples of scenarios yielding negative refraction indices: For the simplest example we set
\beq\label{9}
 \alpha^{\rm 1D} (x) = \delta(x),
\eeq
resembling scatterers at the position $x=0$ with effective field $\xi(0) = 1 = \exp(i|k| 0)$. Further we choose the stationary Green's function for the scatterers to be
\beq\label{ex1}
G^{\rm 1D}_{e1} (x) = \begin{cases} \frac{- i \sin (- |k| |x|)}{ |k|} & \text{ for } x \geq 0\\  \frac{ i \sin (-  |k| |x|)}{ |k|}  & \text{ for } x<0, \end{cases}
\eeq
which can be constructed from the elementary solutions of the Helmholtz equation with point source by taking the superpositions
\begin{equation}
\mp \frac{1}{2} \left( - \frac{ e^{ i |k| |x|}}{ |k|} +\frac{e^{- i |k| |x|}}{ |k|} \right) = \pm \frac{i \sin (- |k| |x|)}{|k|}.
\end{equation}
So the material consists of scatterers inducing an anisotropic wave \eqref{ex1}. The frequency dependent amplitude is set
\beq\label{aut}
f (|k|) = - 2 |k| g(|k|).
\eeq
Thus we have given an example for  $ \alpha^{\rm 1D} (x) = \xi \cdot \rho (x)$ and a condition on the frequency spectrum $f(|k|)$ of the scatterers, which implies a negative refractive material for the generic incoming wave function defined by \eqref{13}. Wave packets can be obtained by integrating \eqref{13} over $k$ with appropriate weighting factors. On the other hand for given $f$ we can model wave functions via $g$ that satisfy \eqref{aut}.

\emph{Example $2$: Anisotropic scatterers.--} The above Green's functions can be generalised to anisotropic scatterers, so that scattering events depend on the direction of the incoming wave. In $1$D the corresponding anisotropic Helmholtz equation for each scatterer is generalised to \cite{berk, berk2}
\beq
C_\pm (x) \partial_x^2 \phi + k^2 \phi =- \delta(x).
\eeq
We assume
\beq
C_\pm (x) = \begin{cases} 1 & \text{ for } \pm x \geq 0\\  -1 & \text{ for } \pm x<0 \end{cases}
\eeq
with solutions
\beq
\phi_\pm (x) = \begin{cases} \pm G^{\rm 1D} (x) & \text{ for } \pm x \geq 0\\  \exp(x |k|) c_1+\exp(-x |k|) c_2 & \text{ for } \pm x<0 \end{cases}
\eeq
where $c_{1,2}$ are chosen to satisfy continuity. Setting $c_1 =0$ we get a decaying solution for $\pm x<0$. For two different species of scatterers positioned at the same points in space (each Green's function approximately being solutions of a Helmholtz equation) we get the superposed effective Green's function
\beq
G^{\rm 1D}_{e2} (x) = \phi_+ (x) +\phi_- (x).
\eeq
Utilizing two species  of scatterers is applicable for high $|k| \gg 1$ as this decreases the relevance of the decaying part of the wave functions and we formally arrive at the previous case. Thus we have generated a similar Green's function as in \eqref{ex1}, but by very different physics involving two different scattering profiles of two species of scatterers occupying the same space.

\emph{Extended scattering domain.--} While the $\delta$ distributed scatterers only resemble an approximation to realistic formations of scatterers, we extend the $1$D analysis to fields,
\beq\label{aaa22}
 \alpha^{\rm 1D} (x) = \frac{1}{\sqrt{4 \pi} \sigma} \exp \left(-\frac{x^2}{\sigma^2} \right) \cdot \exp \left(i k x \right),
\eeq
where the plane wave part is due to the spatial dependence of the effective field $\xi = \exp \left(i k x \right)$ and the remainder models an extended distribution of scatterers.
Hence the l.h.s. \eqref{13} for a stationary Greens function solution \eqref{ex1}  is given by,
\begin{equation}
- \frac{f(k)}{|k|} \int^\infty_{-\infty} \sin (- |k| x) \frac{1}{\sqrt{4 \pi} \sigma} \cdot \exp \left(-\frac{(x-x')^2}{\sigma^2} + i k (x-x') \right) = \frac{f(k)}{|k|} \frac{i  \sqrt{\pi}}{4} e^{- \frac{k^2}{2}  \sigma^2} \sin \left(k \cdot x' + \frac{1}{2} i k^2 \sigma^2 \right).
\end{equation}
Further
\begin{equation}
\sin \left(k \cdot x' + \frac{1}{2} i k^2 \sigma^2 \right) = \sin\left( k \cdot x'  \right)\cosh\left( \frac{1}{2} k^2 \sigma^2\right) + \mathrm{i}\cos\left( k \cdot x' \right)\sinh\left(\frac{1}{2} k^2 \sigma^2\right).
\end{equation}
Again choosing the frequency spectrum $f$ accordingly we can match the r.h.s. in \eqref{13} to obtain negative refraction for $\sigma \to 0$ or for $k \to 0$ or for given $f$ we confine the consideration to appropriately related $g$.

\subsection{Negative refraction in $3$D} Next we turn to the case of negative refraction in $3$D materials.  Green's functions of the $3$D Helmholtz equation for different frequencies $|\kv|$ are
\beq\label{g3d}
 G^{\rm 3D}(\rv) =   \frac{\exp\left(\pm i  |\kv| \cdot |\rv| \right)}{4 \pi |\rv|}
\eeq
with $\rv =(x,y,z) \in \mathbb R^3$, called respectively the outgoing and ingoing spherical waves solutions. Linear combinations of the incoming and outgoing wave solutions for the $3$D Helmholtz equation are e.g. the stationary waves
\beq
 G^{\rm 3D}_{1}(\rv) = \frac{-\cos (|\kv| \cdot |\rv| )}{4 \pi |\rv|}
\eeq
and
\beq\label{an}
 G^{\rm 3D}_2(\rv) = \frac{-i \sin (|\kv| \cdot |\rv| )}{4 \pi |\rv|}.
\eeq
Similar as in the $1$D case we set all scatterers to rest at the origin, but in contrast with a certain spatial distribution given by
\beq\label{aaa1}
 \alpha^{\rm 3D} (\rv) = 4 \pi |\rv| \delta(\rv),
\eeq
while using the  $\xi (\rv =(0,0,0)) =\exp\left(\pm i  \kv \cdot \rv \right) =1$ which neglects all additional multiple scattering processes at this particular scatterer \cite{Foldy}. W.l.og. we set $\kv = (k_x,0,0)$ and in addition we assume the scatterers to induce a Green's function \eqref{an}
\beq\label{ex3}
G^{\rm 3D}_{e3} (\rv) =  \frac{-i \sin (|\kv| \cdot |\rv| )}{4 \pi |\rv|}.
\eeq
Starting from \eqref{sol} we require as condition for negative refraction in $3$D:
\begin{equation}\label{sol22}
 \int_V G^{\rm 3D}(\rv-\rv') f (\kv, \rv')  \rho(\rv') \xi (\rv')  d\rv' \equiv g(\kv)\left( \exp\left(-i \kv \cdot \rv \right) - \exp\left(i \kv \cdot \rv \right) \right) =- 2 i  g(\kv) \sin \left( k_x x \right).
\end{equation}
We assume $f (\kv, \rv') = f (\kv)$.
On the other hand we get using \eqref{aaa1} and \eqref{ex3} 
\begin{equation}
 \int_{\mathbb R^3}  f (\kv) \cdot G^{\rm 3D}(\rv-\rv') \cdot  \alpha^{\rm 3D} (\rv') d \rv'  = - i f (\kv) \sin (|k_x| \cdot |\rv| ).
\end{equation}
Next we apply the paraboloidal wave approximation, $|\rv| \simeq x + \frac{z^2+y^2}{2x}$ that is valid for $\big|\frac{z^2+y^2}{2x} \big| \leq 1$. We are interested in the far field where $x \to \infty$, so that the approximation is satisfied. Thus we obtain
\beq\label{toto}
- i f (\kv) \sin (|k_x| \cdot |\rv| ) \simeq - i f (\kv) \sin (|k_x| \cdot x ).
\eeq
Choosing $f(\kv)$ by comparison of \eqref{toto} with \eqref{sol22} concludes the proof for the far field, i.e. in the $x \to \infty$ limit.

\subsection{Implicit negative refraction}

\emph{Scattering in $1$D.--} While the previous examples provide a general guide to induce negative refraction, we now turn to a specific example feasible in systems satisfying the stated assumptions. For the sake of simplicity we consider the $1$D case. For each $k$ there is a solution $\phi$ with no incoming wave from the right - a standard solution familiar in calculations for reflection and transmission from a generic scattering potential $V(x)$ (with $V = 0$ as $|x| > d$) modelling a single finite scatterer. While here we make specific choices note that the above general framework applies for an ensemble of many (generally randomly distributed) scatterers. Now, we consider an incoming wave 
\beq
\phi_{\rm in} (x) = e^{ikx} + r e^{-ikx} \hspace{12mm} x < -d
\eeq
and consequently we have an outgoing wave
\beq
\phi_{\rm out} (x) = t e^{ikx} \hspace{25mm} x > d,
\eeq
where the reflection probability is $R=|r|^2$ and the transmission probability $T=|t|^2$ - for unitary scattering the conservation of probability applies, i.e.
\beq\label{co}
1- |r|^2 = |t|^2.
\eeq
We assume scatterers to act on finite range, $d < \infty$. Using this generic framework we explicitly construct an example where $\phi_{\rm in} (x,k) = \phi_{\rm out} (x,-k)$ due to multiple scattering events and in contrast to perfect mirrors the solutions are not confined to the area in front of the perfect mirror.

\emph{Refracted waves in $1$D.--} To further elucidate the feasibility of negative refraction in an ensemble of scatterers we proceed as follows. We consider the incoming wave moving from the l.h.s. towards the scatterers which then scatters on the meta-material spanning from $a$ to $c$,
\beq
\phi_{\rm in} (x) = e^{ikx} + r^L_A e^{-ikx} \hspace{12mm} x < a.
\eeq 
The induced wave function between two scatterers, say $A$ and $B$ at positions $a$ and $b$, given there is another scatterer  $C$ at $c$ with $a<b<c$ and subsequent scatterers and scattering events are neglected, is
\begin{equation}\label{formula}
\phi_{A,B} (x) \simeq t^L_A e^{ikx} + t^L_A r^L_B e^{-ikx} + t^L_A r^L_B r^R_A e^{ikx}  + t^L_A t^L_B r^L_C t^R_B e^{-ikx}  \hspace{2mm}  a < x < b,
\end{equation}
where $t^k_i \in \mathbb C$ and $r^k_i \in \mathbb C$ with $i = A,B,C$ and $k=L,R$ are the corresponding transmission or reflection coefficients from the left or right ($L/R$) of the three scatterer system. At the same time each pair of transmission and reflection coefficients has to conserve probability \eqref{co} and if $|r^R_A| <1$ the transmitted wave from the r.h.s. has to be taken into account by consider the effective incoming wave
\beq\label{eff}
\phi_{\rm in} (x,k) \to \phi_{\rm in} (x,k) = e^{ikx} + (r^L_A +t^L_A r^L_B t^R_A) e^{-ikx}.
\eeq 
To get the scattering behavior that induces a negatively refracted wave between $A$ and $B$, i.e. $\phi_{\rm in} (x,k) = \phi_{\rm out} (x,-k) = \phi_{A,B} (x)$, as implied in \eqref{form} we find using \eqref{formula} and \eqref{eff} the necessary conditions
\beq\label{k1}
(r^L_A +t^L_A r^L_B t^R_A) = t^L_A (1+ r^R_A r^L_B)
\eeq
and
\beq\label{k2}
1 = t^L_A (r^L_B + t^R_B t^L_B r^L_C).
\eeq
Note that the setup trivially extends to $N$ scatterers, $A_j$ denoting the $j$th scatterer. By letting the first $A_1$ at position $a_1 \to - \infty$ and the last scatterer $A_N$ at $a_N \to \infty$ we have generated a scenario consistent with \eqref{13} when choosing $t^L_A$ accordingly. Formula \eqref{formula} is an approximation considering only the first and second order reflections and transmissions - iteratively one would obtain the exact form including all reflection and transmission processes.

\emph{Numerical example.--} Let us consider an electron of mass $m = 9.10938291(40) \times 10^{-31} kg$ with velocity $v = 1 m/s$ corresponding to a wave vector \eqref{pv} $k = p/\hbar = 8,7/mm$. Furthermore we consider the three scatterers $A,B,C$ and specify their scattering properties by choosing $t^R_A = \sqrt{3/4}$, $r^R_A = -\sqrt{1/4}$, $r^L_A = 1/2$, $t^L_A = \sqrt{3}/2$, $r^L_B =0.31$, $t^L_B =0.95$, $t^R_B = 1$, $r^L_C =0.89$ to satisfy \eqref{k1} and \eqref{k2}. Then using the above characteristics we obtain the wave functions
\begin{equation}
\phi_{A,B} (x) \simeq 0.73 e^{i 8,7 x}  +  e^{-i 8,7 x}  \equiv  \phi_{\rm out} (x, 8.7) =   \phi_{\rm in} (x, - 8.7)
\end{equation}
where $[x] = mm$ and note that the wave is independent of the actual positions $a,b$ and $c$.




\subsection{Complete tunnelling in negative refraction media} 
\begin{figure}
\vspace{15mm}
\begin{picture}(-50,100)
\put(-125,0) {\includegraphics[width=75mm, height=45mm]{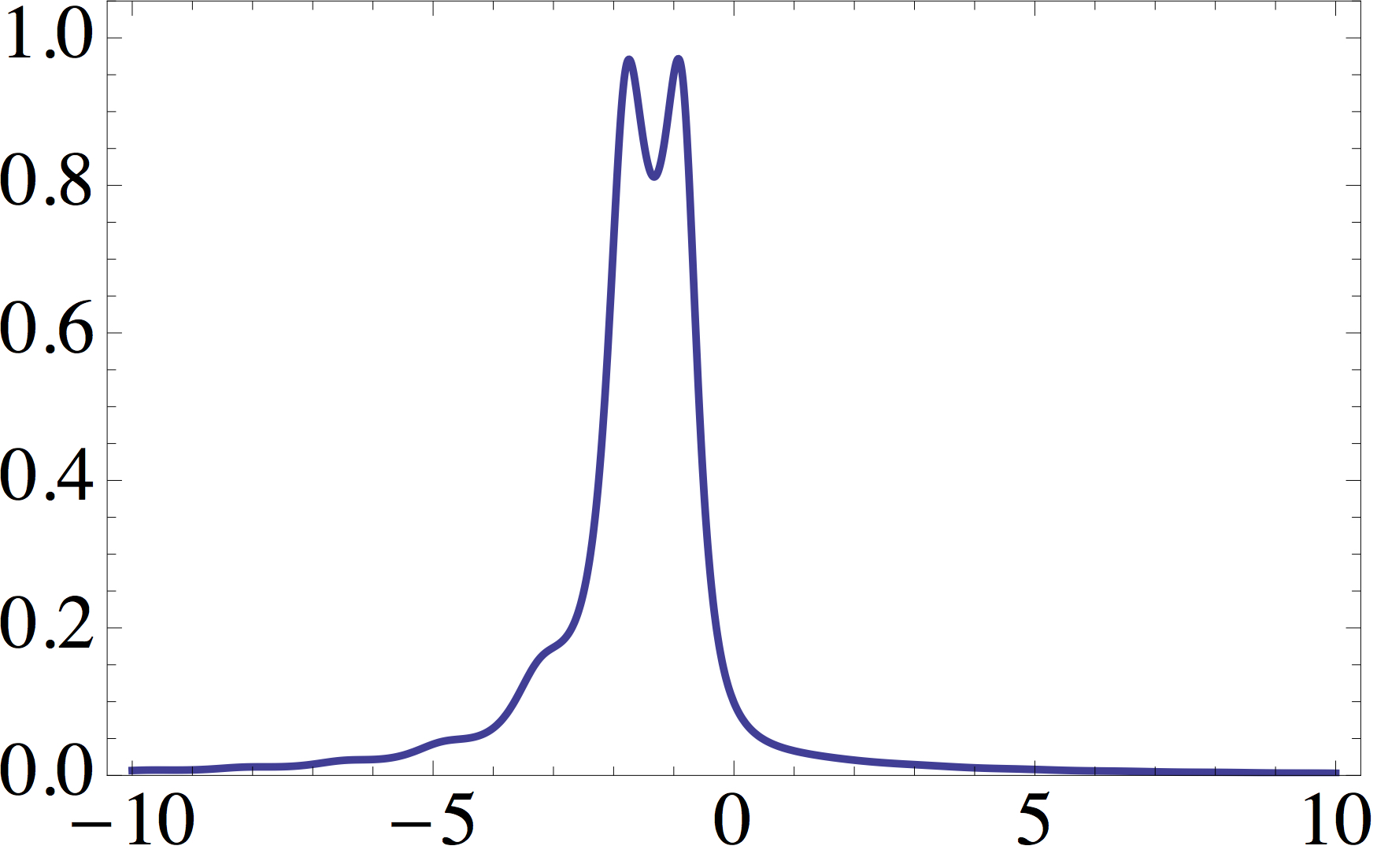} } \put(3,0){\textcolor{black}{${\rm Re}(n)$}} \put(95,100){$T$}
\end{picture}
\vspace{1mm}
\caption{Transmittance $T$ as a function of ${\rm Re}(n)$ for decay $\sigma =0.95$, $c=1.94$ showing amplification of the matter wave for certain negative refraction indices $n$.} \label{trans}
\end{figure}

Next we analyse the transmittance of a matter wave passing a negative refraction index material. 
We assume the material with refraction index $n$ to span from $x=0$ to $x=L$. Therefore the wave function before, within and after the material is
\beq
\psi_{\rm in }(x) = e^{ik x} + r e^{-i k x} \hspace{8mm} x < 0
\eeq
\begin{equation}
\psi_{\rm n < 0}(x) = A e^{\kappa_1 x}+ B e^{- \kappa_1 x}   \hspace{8mm} 0 \leq x \leq L
\end{equation}
\beq
\psi_{\rm out} (x) = t e^{i K x}   \hspace{20mm} L < x
\eeq
correspondingly, where $\kappa_1 = i ( k + \chi + i \sigma )=  i n(k)  k$ is in accordance to \cite{R1,R2,R3}. Here some attenuation $\sigma$ of the wave within the material and a change in phase $\chi$ has been introduced and the transmission amplitude is again $t$. 
At the points $x=0$ and $L$ continuity of the wave function requires
\beq
\begin{cases} 1+r=A+B \\ A e^{ \kappa_1 L} + B e^{ - \kappa_1 L} =  t e^{i k L}
\end{cases}
\eeq
and continuity of its first derivative 
\beq
\begin{cases} i k (1-r) =   \kappa_1  (A-B) \\   \kappa_1 \left( A e^{\kappa_1 L} - B e^{-\kappa_1 L} \right)= i k t e^{i k L}.
\end{cases} 
\eeq
Consequently the transmission amplitude is
\beq
t = \frac{2 n e^{-ik L}}{2 n \cos(k n L) - i (1 + n^2 ) \sin(k n L)}
\eeq
with $\kappa_1 = i n  k$. Further using $n(k) = 1+ \frac{1}{k} (\phi + i \sigma)$ and writing $c=k L$ we obtain the transmittance
\beq
T=|t|^2 = \frac{4 |n|^2}{|2  n \cos(c \cdot n) - i (1 + n^2) \sin(c \cdot n)|^2}.
\eeq
In Fig. \ref{trans} the transmittance is shown as a function of the real part of the refraction index, i.e. ${\rm Re}(n) = 1+ \phi/k$. For the given parameters we observe that a negative refraction medium for matter waves allows their  amplification within the material in analogy to the concept of electromagnetic waves \cite{P1}. However the conservation of probability provides a natural upper bound to the transmittance in a single-body Schr\"odinger equation picture. Recall the relation $1=R+T +A$, where $A$ corresponds to the absorption due to material properties. In case that $T=1$ necessarily $R=A=0$ and we consider the scenario of perfect transmission of a negatively refracted wave through the medium. 
In contrast in positive refraction indices materials usually waves decay - this is particularly shown by the graph of Fig. \ref{trans} for  ${\rm Re}(n) = 1+ \phi/k > 0$. Hence, in this setting complete transmission of a matter wave is feasible only within a negative refraction material. Furthermore it is known that negative refraction of waves allows localisation to a single point given that $n=-1$ as it has been shown in \cite{P1, P2, Turk} for electromagnetic waves, which is a consequence of the modified wave propagation within those meta-materials and applies in one-to-one analogy.   The reason can be understood by the argument that the optical path through a negative refraction material is zero and so the initial point source is mapped on itself. I refer to \cite{P1,P2} and references therein for a complete discussion of this topic.

\section{Discussion}

While negative refraction for electromagnetic waves was regarded a mathematical curiosity until the emergence of meta-materials, it now is an established and well-tested physical phenomenon. The results presented in this paper show the feasibility of negative refraction materials for matter waves due to the variety of free parameters of the many-body problem under consideration namely, the generally complex-valued scattering amplitude, the real-valued number density of local scatterers and the effective local field, which itself is a function of the former two and can be approximated by a plane wave when neglecting multiple scattering processes at each scatterer but considering only the leading order. The general theoretical approach presented here extends to statistical ensembles, but trivially includes the case of a definite choice of an ensemble configuration of scatterers. One striking feature of negative refraction is the possibility of perfect transmission through the medium with $n=-1$ - in contrast to a perfect mirror the outgoing wave occupies the space `behind' the negative refraction material, but with opposite momentum. These features imply the feasibility of static and statistical super-lenses for matter particles with promising applications such as  cloaking of objects regarding influx of specific matter - the object becomes `untouchable' - and perfect focusing of atom beams to a single point. 

In addition to the general framework we have stated an explicit scenario in terms of specific scatterers with realistic numerical values in a formally simpler framework, which provides a first guide for developing meta-materials for negative matter scattering. The stated examples are chosen to be as simple as possible. In the first explicit example the consideration is confined to $1$D and scattering of a single frequency is assumed while wave packets are merely integrals thereof. Furthermore we suppose that the scatterers are located around a single point. The effective field there corresponds to a complex function. Then for the given choice of the incoming wave the associated wave packet is negatively refracted. Extensions thereof show the flexibility of the presented scheme. In addition feasibility has been shown for a $3$D scenario when considering a far field approximation. Limitations of the scheme are clearly given by that fact that only specific particle distributions $\rho$ with a certain fixed frequency dependence of the scattering amplitude $f(x,k)$ and scattering behaviour of their generating Green's function $G$ imply negative scattering. Further when considering statistical ensembles, instead of a definite choice, the considerations made only apply in the large $N$ limit.

\section{Methods}

I utilized multiple scattering theory to analyse the feasibility of negative refraction for matter waves systems due to multiple scattering events. After introducing a general widely accepted and experimentally well-tested approach analytical expressions for the effective local field, Green's functions, density and the frequency behaviour of the scatterers were stated, that consequently yield an outgoing counter-propagating/negatively refracted wave in analogy to the scenario observed for EM waves. The deductive reasoning/mathematical analysis is based on standard analytical tools, e.g. approximations in the large $N$ limit, utilizing continuity properties of the wave function at the surfaces of the negative refractive material - an essential property in quantum mechanics. The latter implied the complete transmission of the quantum particle for particular $n<0$. Finally the proposed negative refractive materials for sub-wavelength localisation of matter waves are an extension from light waves and are justified by analog behaviour of waves in materials with refraction index $n$ independent of their physical nature. 
\[
\]

\section{Acknowledgements} I acknowledge financial support through a Schr\"odinger Fellowship (Austrian Science Fund (FWF): J3675) at the University of Oxford and I am grateful for funding through the NQIT project (EP/M013243/1). I thank Alexander Dreismann (Cavendish Laboratory) for critical discussions and for helping to state the problem. Furthermore I would like to thank the referee for his/her very useful comments.

\section{Additional Information}

There are no competing financial Interests.

  \end{document}